\documentclass[twocolumn,showpacs,preprintnumbers,amsmath,amssymb,floatfix,superscriptaddress]{revtex4}

\usepackage[dvips]{graphicx}
\usepackage{dcolumn}
\usepackage{bm}

\def\beq{\begin{equation}}
\def\eeq{\end{equation}}

\newcommand{\affA}{%
    Van der Waals-Zeeman Institute, University of Amsterdam,\\
Valckenierstraat 65, 1018 XE Amsterdam, The Netherlands }

\newcommand{\affB}{%
    ARC Centre of Excellence for Quantum-Atom Optics and\\
    Centre for Atom Optics and Ultrafast Spectroscopy,\\
    Swinburne University of Technology, Hawthorn, Victoria 3122, Australia}

\begin{document}

\preprint{}

\title{Fully permanent magnet atom chip for Bose-Einstein condensation}

\author{T. Fernholz}
\altaffiliation{Present address: QUANTOP - Danish Quantum Optics
Center, Niels Bohr Institute, 2100 Copenhagen, Denmark}
\email{fernholz@nbi.dk}\affiliation{\affA}
\author{R. Gerritsma}
\altaffiliation{Present address: Institut f\"{u}r Quantenoptik und
Quanteninformation der \"{O}sterreichischen Akademie der
Wissenschaften, Otto-Hittmair-Platz 1, A-6020 Innsbruck, Austria}
\affiliation{\affA}
\author{S. Whitlock}\affiliation{\affA}\affiliation{\affB}
\author{I. Barb}\affiliation{\affA}
\author{R.~J.~C. Spreeuw}\affiliation{\affA}
\affiliation{}\homepage{http://www.science.uva.nl/research/aplp/}

\date{\today}

\begin{abstract}
We describe a proof-of-principle experiment on a fully permanent
magnet atom chip. We study ultracold atoms and produce a
Bose-Einstein condensate (BEC). The magnetic trap is loaded
efficiently by adiabatic transport of a magnetic trap via the
application of uniform external fields. Radio frequency
spectroscopy is used for in-trap analysis and to determine the
temperature of the atomic cloud. The formation of a Bose-Einstein
condensate is observed in time-of-flight images and as a narrow
peak appearing in the radio frequency spectrum.
\end{abstract}



\maketitle

\section{Introduction}
Miniature patterns of magnetic field sources integrated on planar
substrates, `atom chips', are used to manipulate magnetically
trapped clouds of ultracold neutral
atoms~\cite{FolKruHen02,Rei02,DekLeeLor00,ForZim07}. Atom chips
based on current-carrying wires are relatively simple and
versatile tools, readily used to produce Bose-Einstein condensates
(BEC)~\cite{HanHomHan01,OttForSch01}. Subsequently a wide range of
experiments have been performed; for example, BECs can be
precisely positioned with atomic conveyer-belt
potentials~\cite{HanHomHan01,GunKemFor05} and phase-coherent
splitting and interference of condensates has been achieved using
double-well potentials~\cite{SchHofKru05,JoShiPre06}.

Atom chips can also incorporate permanent magnetic materials~
\cite{SinCurHin05,HalWhiSid06,BarGerSpr05,BoyStrPri06,SheHeiPfa06},
with significant additional advantages over current-carrying wires
alone. Specifically, permanent magnet atom chips do not suffer
from current noise which is partly responsible for trap loss and
heating of atoms near conducting surfaces~\cite{SinCurHin05}. They
are also well suited for tightly confining trapping potentials
without the technical problems associated with high current
densities. Furthermore, patterned permanent magnetic materials
provide greater design freedom, allowing complex magnetic
potentials to be realized such as ring-shaped waveguides or large
two-dimensional arrays of
microtraps~\cite{FerGerSpr05,GhaKieHan06,GerWhiSpr07}.

One obvious drawback of permanent magnets however is the
relatively limited degree of dynamic control available over the
associated trapping potentials. The permanent magnetic field is a
particular source of difficulty when loading the atom chip. As
such, most permanent magnetic structures rely on current carrying
wires to load and to complete the permanent magnetic trapping
potential. Analysis is another complication, because standard
techniques such as free ballistic expansion are impossible in the
permanent magnetic field.  These basic challenges need to be
addressed before more sophisticated atom chip designs
incorporating permanent magnets can be realized.

In this paper we describe the loading of ultracold atoms and
evaporative cooling used to produce a BEC in a fully permanent
magnetic microtrap, without additional external control fields.
Although the permanent magnetic chip is in principle
"self-biasing" we use a small field derived from an externally
installed permanent magnet to tune the Ioffe field of the trap. To
load the permanent magnet trap we adiabatically transfer the atoms
to the chip surface from a quadrupole magnetic trap by applying a
combination of bias magnetic fields. During loading we apply a
radio frequency field to evaporatively cool the cloud. Analysis of
the cloud temperature and spectral distribution is then performed
using radio frequency
spectroscopy~\cite{MarHelPri88,HelMarPri92,BloHanEss99,GupHadKet03,ChiBarGri04,WhiHalSid07}.
This technique is also used to observe the BEC transition
\emph{in-situ} without expansion from the trap. It is also
possible to eject the cloud from the self-biased trap by rapidly
switching on uniform magnetic fields to observe the BEC transition
using conventional absorption imaging.

\section{Setup}
\label{secSetup}

\begin{figure}
\includegraphics[width=\columnwidth]{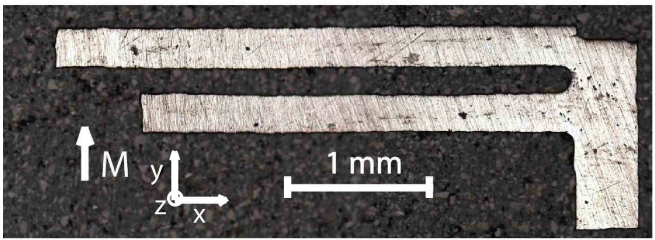}
\caption{Permanent magnetic structure for producing a Ioffe trap
180 $\mu$m above the surface. The structure has a thickness of 40
$\mu$m and an in-plane magnetization of 430 kA/m.}
\label{figStructure}
\end{figure}

Our atom chip consists of a F-shaped permanent magnet structure
(Fig.~\ref{figStructure}), designed to produce a self-biased
Ioffe-Pritchard microtrap 180~$\mu$m from the chip
surface~\cite{BarGerSpr05, XinBarGerPP}.  It is cut from 40 $\mu$m
thick FePt foil by computer numerical controlled (CNC) spark
erosion and is glued to an aluminium-coated glass substrate. The
magnetization is oriented in-plane along the $y$-direction and was
measured using superconducting interference devices
(\textsc{SQUID}) magnetometry as 430~kA/m. Assuming a uniform
magnetization, we calculate an anisotropic, harmonic trapping
potential with trapping frequencies of $2\pi\times$11~kHz and
$2\pi\times$30~Hz in the radial and axial directions respectively.
The calculated magnetic field at the trap minimum (Ioffe field) is
1~G, in the $x$-direction.

It was found experimentally that uncompensated stray fields, such
as the earth magnetic field in combination with field corrugation
caused by the coarseness of the magnetic structure, added field
components in the x-direction on the order of 2 Gauss. The design
aimed at a rather low Ioffe-field to achieve tight radial
confinement, making it necessary to tune the Ioffe-field to a
higher value to avoid field zeros in the trap. This was done with
an externally installed permanent magnet, thus preserving the
permanent magnetic nature of the trap. The permanent magnet was
placed on the optical table ($\approx 25$~cm below the chip),
adding a homogeneous field of $\approx 4$~G in the x-direction.
The trap frequencies in the lowest minimum were then measured to
be $2\pi\times$5.2~kHz radial and $2\pi\times$128~Hz axial with a
field minimum around 2.7~G. Due to the field corrugation, the
axial confinement is significantly larger than expected. We
attribute the field corrugation mainly to the rather
unsophisticated production of the chip and feel that a more
advanced atom chip (for instance, produced with patterning
techniques based on optical lithography) could be fully
self-biased. We note that in fact it was possible to load a
secondary axial minimum of the trap and perform evaporative
cooling without the permanent magnet installed, but less
efficiently.

The atom chip is mounted upside down in a quartz-cell vacuum
chamber ($4\times 4\times 7$~cm$^3$), which is surrounded by three
orthogonal coil pairs that can produce magnetic fields $>100$~G
with any desired polarity. The coils are used to load the atoms
into the permanent magnet trap as described below. The loading
procedure ends by ramping the currents to zero and
short-circuiting the coils with relays. The latter significantly
reduces the residual magnetic field noise. Additional sources of
magnetic field noise, such as the current supplies themselves,
were identified and placed $>5$~m from the experiment. Residual
background field noise at 50 Hz was further reduced by feeding a
phase and amplitude adjusted 50~Hz signal derived from the mains
to a large coil pair placed around the experiment. Our efforts
reduced the root-mean-square magnetic field noise in the $x$
direction from more than 20~mG to below 0.5~mG.  Radio frequency
(rf) fields used for forced evaporative cooling and for rf
spectroscopy of the atom cloud are produced by a computer
controlled 10-bit direct digital synthesis rf generator
\cite{MorBelPer07} (AD9959) with four independent outputs. We use
two outputs connected to separate antennas for evaporative cooling
and rf spectroscopy.

\section{Loading procedure}

\begin{figure}
\includegraphics[width=0.9\columnwidth]{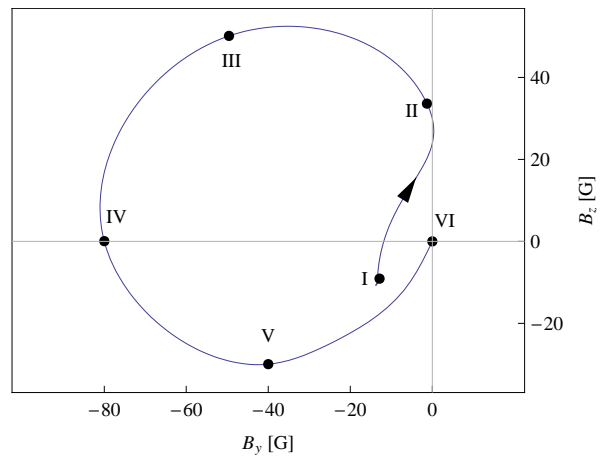}
\caption{Homogeneous magnetic field components applied during the
loading process. Position I corresponds to the initial coil based
quadrupole trap. The magnetic field zero is moved close to the
chip and the anti-Helmholtz coils are subsequently turned off near
position II, turning the trap into a Ioffe trap. This trap is then
moved towards the final location, replacing the initial
self-biased trap. All fields are turned off at position VI.}
\label{figLoadingA}
\end{figure}

\begin{figure}[ht!]
\includegraphics[width=70mm]{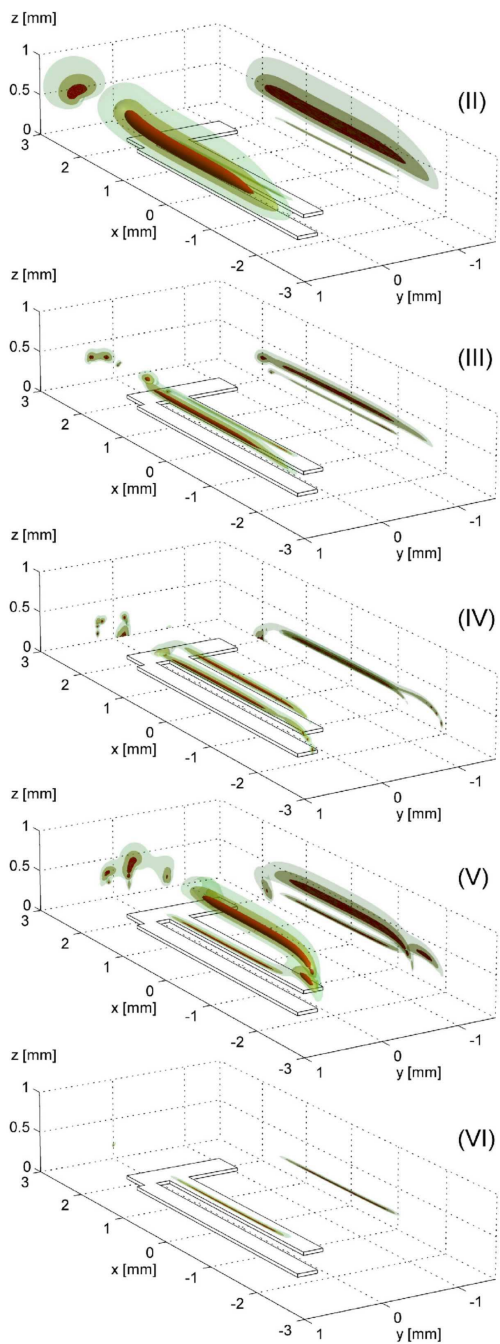}
\caption{(Color online) Simulated contours of magnetic field
strength [corresponding to Fig.~\ref{figLoadingA}(II-VI)] together
with their $x$- and $y$-projections (red=8~G, green=16~G,
blue=24~G) during the loading process. The filled trap on the
left-hand side is compressed and moved towards the chip while the
empty trap is decompressed, moves away from the chip, and
disappears when the field is turned to zero (VI).}
\label{figLoadingB}
\end{figure}

In this section we describe the loading process of the permanent
magnetic trap using external uniform bias fields. The
possibilities and restrictions for moving magnetic traps by this
method have been described in detail in \cite{GerSpr06}.
Efficiently loading atoms into the trap proves to be nontrivial
since at the end of the loading procedure all external fields
should be turned off. The main difficulty arises from the already
present field minimum, located close to the chip, containing no
atoms at the start of the loading process. This minimum can act as
a leak when the trap containing the atoms is brought too close to
it. We have optimized a particular loading trajectory that avoids
this secondary minimum as much as possible.  Another important
feature is that the magnetic field minimum is kept above zero
during the last part of the loading process, making it possible to
evaporatively cool during the transfer without significant loss
due to Majorana spin-flip transitions.

We collect approximately $7\times 10^7$ $^{87}$Rb atoms in a
mirror magneto-optical trap (MOT). The MOT is made with the help
of external field coils, at a distance of 8~mm from the surface
where the field of the chip is negligible. The atomic cloud is
then compressed (CMOT) and optically pumped to the magnetically
trappable $|F=\nobreak2,m_F=\nobreak2\rangle$ state. After
switching all laser beams off, the cloud is trapped 8~mm below the
chip surface using the quadrupole magnetic field produced by one
pair of coils operated in anti-Helmholtz configuration. Additional
uniform fields can bring this trap closer to the chip. In close
proximity to the chip surface the field gradient produced by the
F-shaped magnet is large enough to hold the atoms against gravity
and the anti-Helmholtz coils can be turned off. At this point two
magnetic field minima exist. A weak trap containing the atoms is
produced by the magnet in combination with a uniform bias field
and is located far from the chip surface. A tighter trap close to the chip
surface also originates from the magnet and is at this point still vacant.

\begin{figure}
\includegraphics[width=80mm]{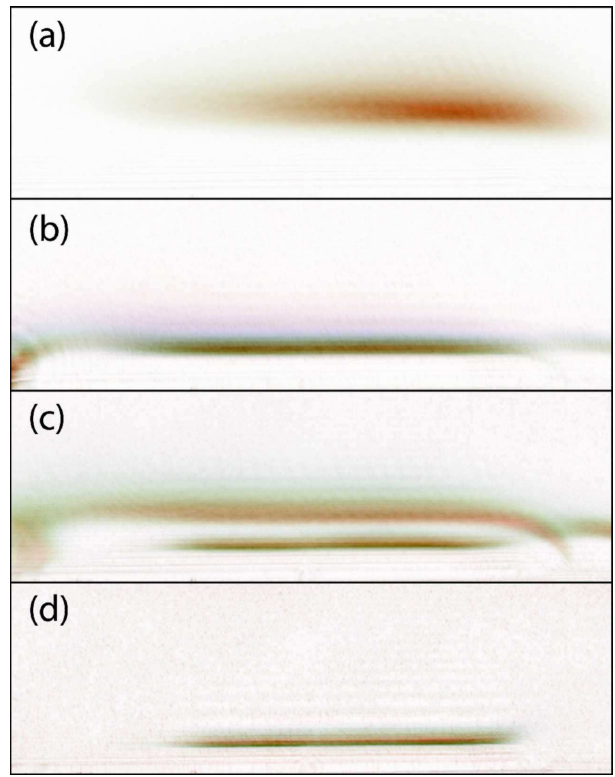}
\caption{(Color online) Spectrally colored absorption images taken
during loading without evaporative cooling. The field of view is
$3.6\times1.7$~mm$^2$. Every image is composed of several
absorption images taken at 22 different probe detunings, ranging
from 0 to 60~MHz. The absorption images are overlapped after
mapping them onto a color code where red corresponds to 0~G Zeeman
shift and blue to $\sim$~30~G Zeeman shift. (a)-(d) correspond to
Fig.~\ref{figLoadingB}~(II,IV-VI). (d) corresponds to the loaded
final trap with all externally controlled fields turned off.}
\label{figLoadspectra}
\end{figure}

An efficient method to transfer atoms to the permanent magnetic
trap is to move the atoms to the location of the final trap while
avoiding the secondary trap. We use a time ordered sequence of
uniform magnetic fields in the $y$- and $z$-directions as shown in
Fig.~\ref{figLoadingA}. The merging of the two minima is prevented
through an additional $B_z$ component of the applied field. The
$B_z$ component is then inverted, as shown in
Fig.~\ref{figLoadingA}, moving the trap containing the atoms
towards the chip, while the empty trap retreats. Finally, after
ramping the fields off, the atom cloud is left in the permanent
magnetic trap.

Three-dimensional calculations of the magnetic field iso-surfaces
during the loading are shown in
Fig.~\ref{figLoadingB}\nobreak(\nobreak II-VI).  During this
transfer the magnetic field gradient of the filled trap increases
from $\sim$~60~G/cm to $\sim$~6800~G/cm which leads to significant
heating of the atom cloud. Although the two field minima do not
merge, they pass each other separated by a finite potential
barrier of $\sim15~$G which allows energetic atoms to spill over
to the decompressing trap. This loss mechanism can take away up to
$~80\%$ of the magnetically trapped atoms. However, it has been
possible to load $2\times10^{6}$ atoms into the permanent magnet
trap.  A series of absorption images at various detunings are
taken during the loading in order to create a series of optical
spectra of the trapped cloud and are shown in
Fig.~\ref{figLoadspectra}. The observed atom distributions compare
very well with the calculated field distributions in
Fig.~\ref{figLoadingB}. The cloud temperature inferred from the
optical spectrum in the self-biased trap is approximately
$T\approx1$~mK, yielding a peak phase space density of
$\rho=N(\hbar\bar{\omega}/k_BT)^3\approx 8\times10^{-7}$. Here,
$N$ is the atom number, $\hbar=h/2\pi$ is Planck's constant,
$\bar{\omega}$ is the geometrical mean trap frequency, and $k_B$
is the Boltzmann factor.

To improve the starting conditions for the final evaporation stage
and to avoid loss due to the finite barrier height we apply forced
radio frequency evaporative cooling already during the compression
of the trap. Evaporative cooling however requires a trap with
non-zero minimum to prevent spin-flip loss. We choose a trajectory
that always maintains a non-zero minimum during the final approach
of the atomic cloud. It is non trivial that such a trajectory
exists~\cite{GerSpr06}. We have found that the trap turns into a
Ioffe trap after the turn off of the anti-Helmholtz coils and
remains a Ioffe trap during the rest of the loading process (steps
II-VI in Fig.~\ref{figLoadingB}). The preliminary evaporative
cooling stage during the transfer consists of a 6~s linear rf ramp
from 35~MHz to 8~MHz. At the end of the ramp all externally
controlled fields are turned to zero and the cloud is trapped in
the potential produced by the magnetic structures alone. After
loading, the cloud contains approximately $10^5$ atoms at a
temperature of $60~\mu$K yielding a peak phase space density of
$\rho\approx 2\times10^{-4}$. Although the number of atoms is
relatively small, the high trap frequencies result in a mean
collision rate of $\sim200 ~$s$^{-1}$. A final rf evaporative
cooling stage is performed with a linear rf sweep over 2.3~s from
8~MHz down to $\sim$1.9~MHz which lowers the trap depth from
600~$\mu$K to $10~\mu$K.

\section{Radio frequency spectroscopy}

\begin{figure}
\includegraphics[bb=130 0 330 190,clip=, width=\columnwidth]{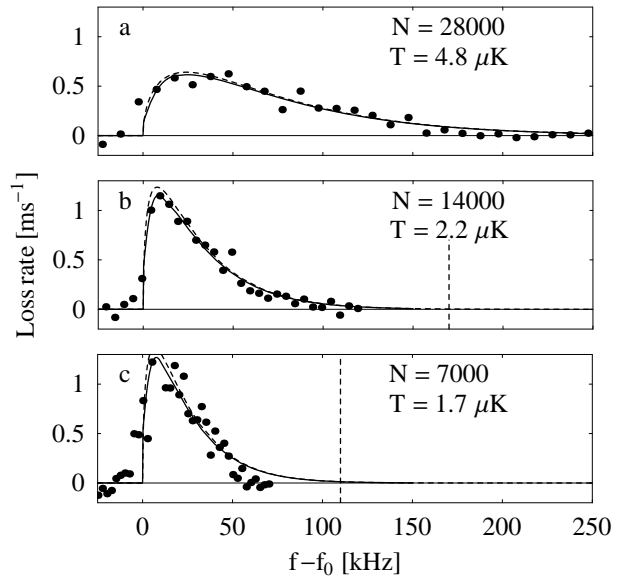}
\caption{Radio frequency spectra taken for various final
evaporation frequencies of 370, 170, and 110~kHz above the trap
bottom $f_0$, as indicated by the dashed vertical lines [(a)-(c),
respectively]. The model used to describe the spectra is obtained
by numerical integration of the Landau-Zener spin-flip probability
assuming Bose-Einstein statistics (solid lines). The calculations
consider the anisotropic trap geometry and give a Rabi frequency
of $\Omega\approx2\pi\times 0.8$~kHz. The dashed lines indicate
the prediction of the potential energy model using
Eq.~(\ref{eqSpecdens}) with the same parameters.
}\label{figspecs}
\end{figure}

The most commonly used diagnostic technique for ultracold atoms is
free ballistic expansion followed by absorption imaging. In our
case this is not possible due to the permanent character of the
magnetization of our chip. Instead, we employ radio frequency
spectroscopy to characterize the ultracold atom cloud in the
permanent magnetic trap. This allows us to directly probe the
energy distribution of the confined cloud and to extract the cloud
temperature without ballistic expansion.

The cloud is probed by applying a radio frequency pulse of
frequency $f$ for a duration of 0.5~ms. Trapped atoms satisfying
the resonant condition $hf=g_F\mu_BB(x,y,z)$ may be coupled to
untrapped magnetic states and are then immediately lost from the
trap. Here, $g_F$ is the atomic Land\'{e} factor, $\mu_B$ is the
Bohr magneton, and $B$ is the local magnetic field modulus. We
measure the number of lost atoms by applying a magnetic field of
25~G in the $y$ direction to push the atoms away from the chip
surface and to weaken the trapping potential for detection.
Resonant absorption imaging can then provide an accurate measure
of the number of remaining atoms.

In Fig.~\ref{figspecs} radio frequency spectra are shown for
various final evaporation frequencies. Above the BEC transition
temperature $T_c$, we can fit a theoretical model to the radio
frequency spectra to obtain the temperature of the cloud, whereas
below $T_c$ radio frequency spectroscopy can be used to observe
the BEC transition without free expansion and to measure the
chemical potential $\mu$. The following sections describe both
techniques.

\subsection{Thermometry}

\begin{figure}
\includegraphics[width=80mm]{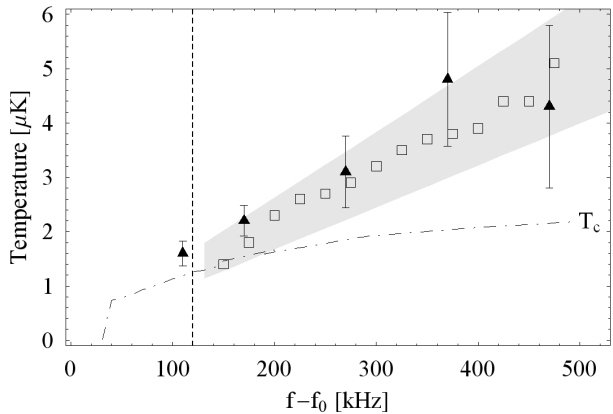}\\
\caption{Temperature inferred from the axial density distribution
($\square$) compared to temperature obtained from rf spectra
($\blacktriangle$) as a function of final evaporation radio
frequency with respect to the trap bottom. The gray area gives the
confidence interval for the temperature inferred from the axial
density distribution, based on estimated systematic errors on the
pixel size and lensing errors. The curve for $T_c$ is obtained
from the atom number and the trap frequencies, and is corrected
for the finite size of the cloud \cite{KetDru96}. The dashed line
at 120 kHz corresponds to the observed onset of condensation.}
\label{figaxsize}
\end{figure}

If we assume that the atoms move through the resonant surface at a
constant speed perpendicular to the surface, we can describe the
spin-flip transition probability by the Landau-Zener expression
\cite{VitSuo97,Wit05}. To obtain a model for the loss rate
spectrum we integrate the spin-flip probability over the velocity
distribution and over the resonant surface. We introduce the radio
frequency above the trap bottom $\omega_{\mathrm{rf}}=2\pi(f-f_0)$
and the thermal frequency $\omega_{\mathrm{th}}=k_BT/\hbar$. In
the case of a gas well above the condensation temperature $T \gg
T_c$ we find that the loss rate in an isotropic trap is well
described by
\begin{align}
\frac{dN}{dt}\approx\left(\frac{1}{2\pi\Omega^{2}}+\frac{\sqrt{m_F}}
{4\bar{\omega}\sqrt{\pi\omega_{\mathrm{th}}\omega_{\mathrm{rf}}}}\right)^{-1}
n_{\omega_{\mathrm{rf}}},
\label{eqSpec}
\end{align}
where we introduced the resonant atomic spectral distribution
$n_{\omega_{\mathrm{rf}}}=dN/d\omega_{\mathrm{rf}}$, using
Boltzmann statistics:
\begin{align}
n_{\omega_{\mathrm{rf}}}=\frac{2N}{\sqrt{\pi}}\sqrt{\frac{m_F^3\omega_{\mathrm{rf}}}
{\omega_{\mathrm{th}}^3}}
e^{-\frac{m_F\omega_{\mathrm{rf}}}{\omega_{\mathrm{th}}}}.
\label{eqBoltzmann}
\end{align}

%
%

As it can be seen from Eq.~(\ref{eqSpec}), the loss rate becomes
independent of the trapping parameters and reflects the potential
energy distribution [Eq.~(\ref{eqBoltzmann})] for sufficiently
small Rabi frequencies
$\Omega^2\ll\bar{\omega}\omega_{\mathrm{th}}/\sqrt{m_F}$. This
arises from the fact that the out-coupling rate becomes
independent of the momentum distribution under the above
assumptions.

For temperatures approaching $T_c$, Eq.~(\ref{eqBoltzmann}) is no
longer valid since it is based on Boltzmann statistics. For
Bose-Einstein statistics we cannot evaluate the integrals to
obtain an expression for $dN/dt$ similar to Eq.~(\ref{eqSpec}).
Instead we integrate the Landau-Zener probability numerically,
including the trap anisotropy. To obtain the cloud temperature
from our data, we fix $\mu$ using the measured atom number $N$ by
setting $\int n_{\omega_{\mathrm{rf}}}d\omega_{\mathrm{rf}}=N$.
The results of this calculation are shown as solid lines in
Fig.~\ref{figspecs}.

We can alternatively assume that Eq.~(\ref{eqSpec}) also holds for
Bose-Einstein statistics with the correct potential energy
distribution (dashed lines in Fig.~\ref{figspecs}). For a
non-interacting cloud with $T \gtrsim T_c$ the resonant spectral
distribution is given by
\begin{align}
\label{eqSpecdens} n_{\omega_{\mathrm{rf}}}=4\frac{\sqrt{\pi
m_F^3\omega_{\mathrm{th}}^3\omega_{\mathrm{rf}}}}{\bar{\omega}^3}g_{3/2}\left(e^{\frac{\mu/\hbar-m_F\omega_{\mathrm{\mathrm{rf}}}}{\omega_{\mathrm{th}}}}\right)
\end{align}
Here $g_{3/2}(x)$ is the polylogarithm function with base 3/2, and
$\mu$ is the chemical potential.  The inferred temperatures
deviate by less than 15\% from the numerically obtained results.

As an alternative temperature measurement we also observe the
axial density distribution of the cloud in the trap. To avoid
problems in the absorption imaging due to the high optical density
of the cloud we lift the magnetic field at the trap bottom by 50 G
to weaken the radial confinement, and detune the probe laser by
18~MHz to the blue of the $\sigma^+$ transition. By fitting the
axial density distribution we obtain the temperature of the cloud.

In Fig.~\ref{figaxsize} we compare the temperatures found from
spectra and the inferred temperature from the axial density
distribution of the cloud for different final evaporation radio
frequencies. Also shown is the confidence interval of the
temperature based on the axial size. This interval is estimated
for systematic errors such as lensing of the probe light by the
dense cloud, pixel size calibration errors, and finite imaging
resolution. The plot shows excellent agreement between the two
temperature measurements.

\subsection{Bose-Einstein condensation}
\begin{figure}
\includegraphics[width=\columnwidth]{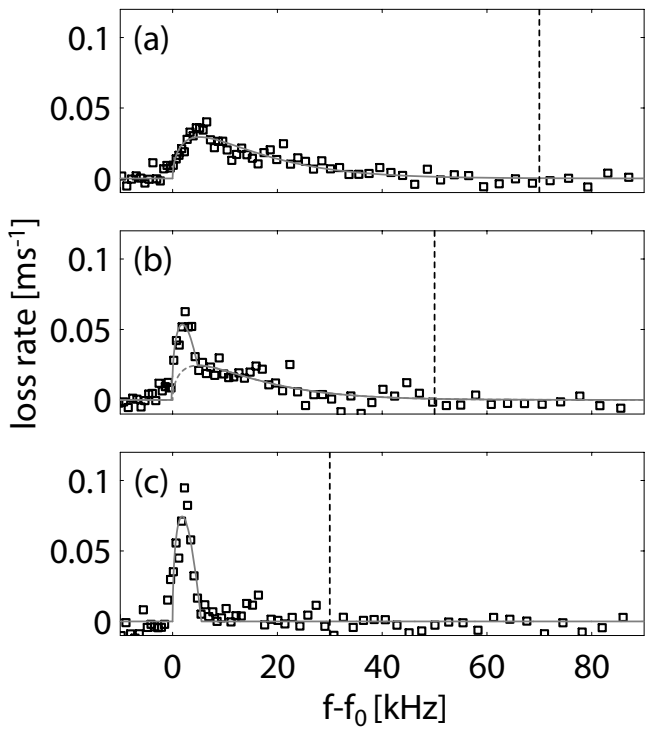}\\
\caption{Spectra of (a) a cloud close to $T_c$, (b) a partially
condensed cloud, and (c) a nearly pure condensate. The dashed
lines indicate the position of the final radio frequency cooling
``knife'' with respect to the trap bottom ($f_0\approx 1.85$~MHz).
The full width of the condensate spectrum is $\sim$~5~kHz,
corresponding to the chemical potential $\mu/m_Fh$. The bimodal
fits, shown in light gray solid and dashed lines for the
condensate and thermal part, respectively, are based on a
heuristic model. } \label{figBECspectra}
\end{figure}

To observe the onset of Bose-Einstein condensation we increased
the probing time to 10~ms and decreased the radio frequency
amplitude by a factor of 3 to minimize power broadening. It took
considerable effort in magnetic field noise reduction to obtain
the resolution needed to observe the narrow condensation peak. In
addition, it was necessary to maintain the same cycle for the
applied magnetic fields throughout the experiment to minimize
hysteresis effects in the magnetic material and surrounding
structures, which otherwise could lead to variations in the trap
bottom of several $10 $~kHz. The observed spectra below $T_c$ are
shown in Fig.~\ref{figBECspectra}. Lowering the final radio
frequency results in a bimodal spectrum and eventually in a
spectrum of a nearly pure condensate with a width of $\sim$~5~kHz.
This agrees well with the chemical potential derived from atom
number and trap frequencies of $\mu/{m_F h}$ = 4.3~kHz. Each
spectrum is averaged 6 times and consists of about 70 data points,
spread nonlinearly over the spectral range to emphasize the
bimodal nature and condensate peak.

\section{Expansion}

\begin{figure}
\includegraphics[width=85mm]{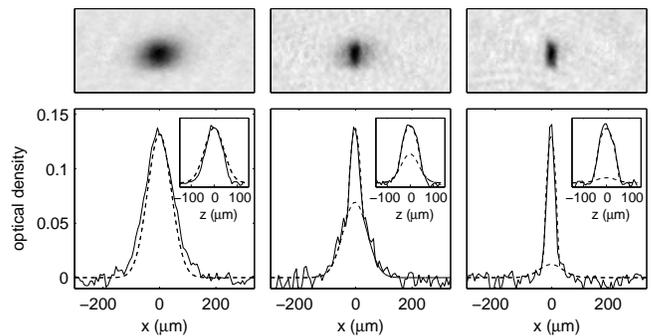}\\
\caption{Absorption images taken after expansion of the cloud (10
times averaged). The atoms are first taken out of the self-biased
trap by field ramps. Then the fields are suddenly turned off and
the cloud falls and expands for 3.5~ms. From left to right, the
figures show a thermal cloud, a partly condensed cloud, and a
nearly pure condensate of 1500 atoms, corresponding to final rf
frequencies of 1970, 1920, and 1890~kHz. The lower graphs and
insets show axial and radial cross sections, respectively, through
the center of the cloud, with two-dimensional bimodal fits shown
with dashed lines.} \label{figBEC}
\end{figure}

As an alternative way to identify the BEC transition, we ramp on,
over 2~ms, a magnetic field of 50~G, which rapidly transfers the
atom cloud 1~mm away from the chip surface. At this position all
magnetic fields are suddenly switched off and the acquired kinetic
energy of the atoms is sufficient to overcome the remaining
potential barrier. The cloud is accelerated further by gravity and
the rapidly decaying field of the chip. After 3.5~ms an absorption
image is taken to reveal a partially expanded atom cloud
(Fig.~\ref{figBEC}). As the final trap depth is reduced, the atom
distribution becomes bimodal with a narrow central peak and a
change in aspect ratio. Combining these absorption images with the
radio frequency spectra we find the BEC transition at a trap depth
of about $11~\mu$K, with 7000 atoms at a temperature of
$T_c\sim~1.6~\mu$K, which is close to the expected value of
$T_c=1.3~\mu$K. Continued evaporation results in a nearly pure BEC
consisting of $\sim 1500$ atoms.

Although this ejection method is useful for observing the phase
transition to BEC it cannot be reliably used to obtain
temperatures because of the non adiabatic nature of the expansion
and the large magnetic field gradients during the expansion. It
can however be used to obtain condensate fractions.

We observe strong heating of the ultracold cloud
($\sim$~6~$\mu$K/s) during the final stage of evaporation. This is
likely caused by three-body collisions as the estimated density at
the trap center near $T_c$ is quite high, $n(0)\approx 6 \times
10^{14}$~cm$^{-3}$. As a result, the lifetime of the condensate is
$\sim$~15~ms without a rf shield and $\sim$~150~ms with a rf
shield \cite{MewAndDru96,BurGhrMya97}. The quoted density
corresponds to a three-body decay rate of $\sim$~3~s$^{-1}$
\cite{BurGhrMya97,SodGueDal99}. It is thus likely that three-body
decay limits the final number of condensate atoms during the last
evaporation stage.

It is interesting to note that condensation is observed in the
expanded cloud at higher evaporation frequencies than in the rf
spectra. We suspect this is because of the long rf probing time
necessary to obtain the desired resolution and signal-to-noise
ratio, causing considerable heating during the radio frequency
pulse. It is also possible that the signal-to-noise ratio does not
allow the observation of small condensate fractions, because it
requires the removal of a substantial fraction of atoms.

\section{Conclusions}
In conclusion, we have presented solutions to the problems of
loading and analysis of an atomic cloud on a fully permanent
magnetic chip. We have demonstrated an efficient loading procedure
relying on the adiabatic movement of Ioffe traps using uniform
external fields, allowing the application of forced evaporative
cooling during loading. After loading, radio frequency
spectroscopy is used to characterize the cloud. A model is derived
for fitting rf spectra to obtain cloud temperatures. The fitted
temperatures compare well to temperatures derived from fitting the
in-trap axial density distribution. Noise reduction and radio
frequency power reduction give the resolution needed to observe
the onset of Bose-Einstein condensation in radio frequency
spectra. This results in bimodal spectra and eventually in a
spectrum of a nearly pure condensate. Alternatively, condensation
is observed by ejecting the atomic cloud by switching external
magnetic fields. Together, these techniques have allowed us to
analyze our atomic cloud in the permanent magnetic field.

\begin{acknowledgments}
We gratefully acknowledge Y.~T. Xing and J.~B. Goedkoop for the
production of the magnetic structure. This work is part of the
research program of the Stichting voor Fundamenteel Onderzoek van
de Materie (Foundation for the Fundamental Research on Matter) and
was made possible by financial support from the Nederlandse
Organisatie voor Wetenschappelijk Onderzoek (Netherlands
Organization for the Advancement of Research). This work was also
supported by the EU under contract MRTN-CT-2003-505032.

\end{acknowledgments}


\begin{thebibliography}{10}

\bibitem{FolKruHen02}
R. Folman {\it et~al.}, Adv.\ At.\ Mol.\ Opt.\ Phys.\ {\bf 48},
263  (2002).

\bibitem{Rei02}
J. Reichel, Appl.\ Phys.\ B {\bf 75},  469  (2002).

\bibitem{DekLeeLor00}
N. Dekker {\it et~al.}, Phys.\ Rev.\ Lett.\ {\bf 84},  1124
(2000).

\bibitem{ForZim07}
J. Fortagh and C. Zimmermann, Rev.\ Mod.\ Phys.\ {\bf 79},  235
(2007).

\bibitem{HanHomHan01}
W. H{\"a}nsel, P. Hommelhoff, T.~W. H{\"a}nsch, and J. Reichel,
Nature {\bf
  413},  498  (2001).

\bibitem{OttForSch01}
H. Ott {\it et~al.}, Phys.\ Rev.\ Lett.\ {\bf 87},  230401
(2001).

\bibitem{GunKemFor05}
A. G{\"u}nther {\it et~al.}, Phys.\ Rev.~A {\bf 71},  063619
(2005).

\bibitem{SchHofKru05}
T. Schumm {\it et~al.}, Nature Physics {\bf 1},  57  (2005).

\bibitem{JoShiPre06}
G.-B. Jo {\it et~al.}, Phys. Rev. Lett. 98, 030407 (2007).

\bibitem{SinCurHin05}
C.~D.~J. Sinclair {\it et~al.}, Phys.\ Rev.~A {\bf 72},  031603(R)
(2005).

\bibitem{HalWhiSid06}
B.~V. Hall {\it et~al.}, J.\ Phys.~B: At.\ Mol.\ Opt.\ Phys.\ {\bf
39},  27
  (2006).

\bibitem{BarGerSpr05}
I. Barb {\it et~al.}, Eur.\ Phys.\ J.\ D {\bf 35},  75  (2005).

\bibitem{BoyStrPri06}
M. Boyd {\it et~al.}, Phys.\ Rev.\ A {\bf 76}, 043624 (2007).

\bibitem{SheHeiPfa06}
A. Shevchenko {\it et~al.}, Phys.\ Rev.~A {\bf 73},  051401(R)
(2006).

\bibitem{FerGerSpr05}
T. Fernholz, R. Gerritsma, P. Kr{\"u}ger, and R.~J.~C. Spreeuw,
Phys. Rev. A {\bf 75}, 063406 (2007).

\bibitem{GhaKieHan06}
S. Ghanbari, T.~D. Kieu, A. Sidorov, and P. Hannaford, J.~Phys.~B:
At. Mol. Opt. Phys. {\bf 39},  847  (2006).

\bibitem{GerWhiSpr07}
R. Gerritsma {\it et~al.}, Phys.\ Rev.~A {\bf 76},  033408 (2007).

\bibitem{MarHelPri88}
A.~G. Martin {\it et~al.}, Phys.\ Rev.\ Lett.\ {\bf 61},  2431
(1988).

\bibitem{HelMarPri92}
K. Helmerson, A.~G. Martin, and D.~E. Pritchard, J.\ Opt.\ Soc.\
Am.~B {\bf 9},
   483  (1992).

\bibitem{BloHanEss99}
I. Bloch, T.~W. H{\"a}nsch, and T. Esslinger, Phys.\ Rev.\ Lett.\
{\bf 82},
  3008  (1999).

\bibitem{GupHadKet03}
S. Gupta {\it et~al.}, Science {\bf 300},  1723  (2003).

\bibitem{ChiBarGri04}
C. Chin {\it et~al.}, Science {\bf 305},  1128  (2004).

\bibitem{WhiHalSid07}
S. Whitlock {\it et~al.}, Phys.\ Rev.~A {\bf 75},  043602  (2007).

\bibitem{XinBarGerPP}
Y.~T. Xing {\it et~al.}, J. Magn. Magn. Mater. {\bf 313}, 192 (2007).

\bibitem{MorBelPer07}
O. Morizot {\it et~al.}, arXiv:0704.1974  (2007).


\bibitem{GerSpr06}
R. Gerritsma and R.~J.~C. Spreeuw, Phys.\ Rev.~A {\bf 74},  043405
(2006).

\bibitem{VitSuo97}
N.~V. Vitanov and K.-A. Suominen, Phys.\ Rev.~A {\bf 56},  R4377
(1997).

\bibitem{Wit05}
C. Wittig, J. Phys. Chem. B {\bf 109},  8428  (2005).

\bibitem{KetDru96}
W. Ketterle and N. {van Druten}, Phys.\ Rev.~A {\bf 54},  656
(1996).

\bibitem{MewAndDru96}
M. Mewes {\it et~al.}, Phys.\ Rev.\ Lett.\ {\bf 77},  416  (1996).

\bibitem{BurGhrMya97}
E.~A. Burt {\it et~al.}, Phys.\ Rev.\ Lett.\ {\bf 79},  337
(1997).

\bibitem{SodGueDal99}
S\"{o}ding {\it et~al.}, Appl.\ Phys.\ B {\bf 69}, 257, (1999).

\end{thebibliography}
\end{document}